\documentstyle[amsfonts,aps,prd,epsf]{revtex}

\newcommand{\beq}{\begin{equation}}
\newcommand{\eeq}{\end{equation}}
\newcommand{\beqa}{\begin{eqnarray}}
\newcommand{\eeqa}{\end{eqnarray}}
\newcommand{\non}{\nonumber}

\begin{document}

\draft

\title{Quenched complexity of the p-spin spherical spin-glass with 
external magnetic field}

\author{
Andrea Cavagna\cite{thanksAC},
Juan P. Garrahan\cite{thanksJPG}
and Irene Giardina\cite{thanksIG}
}

\address{
Theoretical Physics, University of Oxford \\
1 Keble Road, Oxford, OX1 3NP, UK.
}

\date{July 15, 1998}

\maketitle

\begin{abstract}
We consider the $p$-spin spherical spin-glass model in the presence 
of an external magnetic field as a general example of a mean-field 
system where a one
step replica symmetry breaking (1-RSB) occurs. In this context we compute
the complexity of the Thouless-Anderson-Palmer states, performing  
a quenched computation. We find what is the
general connection between this method and the standard static 1-RSB one, 
formulating a clear mapping between the parameters used in the two different 
calculations. We also perform a dynamical analysis of the model, by which
we confirm the validity of our results.
\end{abstract}


\section{Introduction}

Two different sets of results can be obtained when studying a
statistical model. The first set is given by a static (thermodynamical)
analysis of the system, while the second one is obtained from a
dynamical approach (eg., Langevin dynamics).  In the case of disordered
models as spin-glasses, these two sets of results are not in general
connected in a straightforward way.  The reason for this is that
disordered models always exhibit an out of equilibrium dynamical
behavior, which is believed to be related to the presence of many
metastable states.

However, there is a striking correspondence between certain types of
dynamical scenarios and their specific static counterparts.  In other
words, some particular dynamical results are always associated with the
same set of static results and vice-versa. For instance, in the context
of the static replica symmetry breaking (RSB) 
scheme for mean-field spin glasses \cite{spin}, we always find a
dynamical asymptotic energy larger than the static one in models
which are solved by a one step solution (1-RSB), while the two energies 
are the same in models statically solved by a full
RSB solution \cite{crisa1,ck1,ledu,sk,ck2}.  
This fact suggests that there must be some
kind of underlying general physical explanation for these connections.

There is a third method which can be used in the study of disordered
systems, which has the advantage to catch some of the results both of
the statics {\it and} of the dynamics. This method is purely entropic, 
that is it just deals with the number of states of the system,
disregarding their thermodynamical Boltzmann weights.  
In this context, one simply computes the number of local minima
of some mean-field free energy, function of the local magnetizations
of the system, and obtains in this way the spectrum of the states
as a function of their free energy density \cite{tap,crisatap,kpz}.  
This approach gives information both on the 
statics and the dynamics of the system since it deals with the {\it whole}
underlying free energy landscape. We believe that in this context it
should be possible to find an explanation for the relation we stated
above between statics and dynamics.

As a first step in this direction it is necessary to understand how
general and feasible this entropic approach is. 
Up to now it has given correct
results only in one specific mean-field
model with 1-RSB solution, the $p$-spin
spherical spin-glass in zero magnetic field 
\cite{crisa1,kpz,crisatap}. In this
model there is a band of free energy densities where the number of
states i sexponentially large in the size of the system.  It has been
found that the static and the dynamic free energies 
(the last being defined as 
the free energy of the states reached by the dynamics)
are equal to the lowest and highest edges of this band, 
respectively.  We believe this to be true in {\it any} 1-RSB model. Yet, 
in order to find out how general this
result is, it is not sufficient to consider 
the $p$-spin model in zero field.
Indeed, due to the absence of the field, this model has some
particular features, which make it a special case 
in the 1-RSB class. It is homogeneous and therefore presents no
chaoticity of the states with respect to the temperature \cite{kpz}.  
The static and dynamical transitions are both discontinuous, 
which is not a general property of 1-RSB models.
Finally, the dynamical off-equilibrium behavior of the $p$-spin model
with zero field turns 
out to be much simpler than in the general case \cite{ledu}.  
However, this model can be easily
generalized by the introduction of an external magnetic field. It then
becomes a very general paradigm of 1-RSB models, since as a function of
temperature and field it displays all the possible characteristics of 
these systems \cite{crisa1}.

In the present 
paper we extend the entropic calculation mentioned above to the
$p$-spin spherical model in presence of a field. 
In this context we are able to state
in the most general way which are the connections with the standard
static and dynamical approaches.  In particular, we find a one to one
mapping between the 1-RSB static parameters and the ones of the
entropic calculation, giving a physical interpretation of this
relation, until now rather unclear.

The paper is organized in the following way. In Section \ref{sec:model}
we define the model and compute the average logarithm of the number of
states, known as {\it complexity}.  In the next Section we discuss the
results and find the relation with the static 1-RSB calculation. In
Section \ref{sec:plane} we study the behavior of the complexity as a
function of temperature and magnetic field. In Section
\ref{sec:dynamics} we perform a dynamical analysis of the model and
compare it with the results coming from the complexity.  Finally, we
state our conclusions in Section \ref{sec:conclusions}.

\section{Model and Complexity} \label{sec:model}

The $p$-spin spherical model is defined by the Hamiltonian, 
\beq
H(s) = - \sum_{i_1 < \dots < i_p} 
	J_{i_1 \dots i_p} s_{i_1} \dots s_{i_p}
	- h \sum_{i} s_i .
	\label{mod}
\eeq
The spins $s$ are real variables satisfying the spherical constraint
$\sum_i s_i^2 = N$, where $N$ is the size of the system. The couplings
$J$ are Gaussian variables with zero mean and variance $p!/2N^{p-1}$
and $h$ is an external magnetic field \cite{grome,tirumma,crisa1}.  In
the context of the Thouless-Anderson-Palmer (TAP) approach \cite{tap},
it is possible to formulate a set of mean field equations for the local
magnetizations $m_i = \langle s_i \rangle$.  A mean-field
free energy density $f_{\rm TAP}$, function of the magnetizations
$m_i$, was introduced for this model in \cite{kpz},
\beq
f_{\rm TAP} = - \frac{1}{N} \sum_{i_1 < \dots < i_p}
	J_{i_1 \dots i_p} m_{i_1} \dots m_{i_p} 
	- \frac{h}{N} \, 
	\sum_i m_i - \frac{1}{2 \beta} \log{(1-q)} + g(q) ,
	\label{ftap}
\eeq
where $q = N^{-1} \sum_i m_i^2$ is the self-overlap related to the
magnetization $m$ and $g(q)= - \frac{\beta}{4} \left[ (p-1)q^p-p
q^{p-1}+1 \right]$ is the Onsager reaction term (for a derivation of
the TAP free energy see also \cite{rieger,crisatap}).  The
minimization of (\ref{ftap}) with respect to the $\{m_i\}$ gives the
TAP equations of the system,
\beq
{\cal T}_k(m) \equiv 
	- p \sum_{i_2 < \dots < i_p} J_{k, i_2 \dots i_p}
	m_{i_2} \dots m_{i_p} - h 
	+ 2 \, m_k \, 
	\left( \frac{1}{2 \beta \, (1-q)} + g'(q) \right)
	= 0  .
	\label{tap}
\eeq
In the low temperature phase these equations admit many possible
solutions, corresponding to different stationary points of the mean
field free energy (\ref{ftap}). The minima among all the stationary
points can be identified with stable and metastable states of the
system.

The {\em complexity} $\Sigma(f)$ is defined in the following way,
\beq
\Sigma(f) \equiv \lim_{N \rightarrow \infty} \frac{1}{N}
	\ \overline{\log{{\cal N}(f)}} , 
	\label{she}
\eeq
where ${\cal N}(f)$ is the number of local {\it minima} of the free energy
(\ref{ftap}), that is, the number of states of the system with free
energy density $f$. We average the logarithm of $\cal N$ since we
expect this to be the extensive quantity.  
In order to perform this average it is necessary to
introduce replicas. However, when the external field $h$ is set equal 
to zero, the correct ansatz for the overlap matrix turns out to be
symmetric and diagonal. This is equivalent to averaging directly the
number ${\cal N}$ of solutions (annealed average)
\cite{crisatap}. The physical reason for this is that when $h=0$ there
is no preferred direction in the phase space sphere, thus the
typical states are orthogonal to each other and their mutual overlap is
zero. On the other hand, when $h \neq 0$ there is a migration of the
states towards the direction of the magnetic field and their mutual
overlap is different from zero. In this case the quenched average has
to be performed. Anyhow, as it is shown below, due to the particular
nature of the calculation, the correct ansatz for the overlap matrix is
simpler than the 1-RSB used in the statics \cite{crisa1}.

In order to compute $\Sigma$ we use the replica trick,
\beq
\overline{\log {\cal N}} =
	\lim_{n \rightarrow 0} \frac{1}{n}
	\log \overline{ \prod_{a=1}^{n}{\cal N}^{a} }  ,
	\label{trick}
\eeq
where each ${\cal N}^a$ is given by
\beq
{\cal N}^a(f) = \int {\cal  D} m^a  
	\prod_{k=1}^N \delta \left( {\cal T}_k(m^a) \right)  
	\left| \det{{\cal A}(m^a)} \right| 
	\delta \left( f_{\rm TAP}(m^a) - f \right) ,
	\label{number}
\eeq
and ${\cal A}_{kl}(m) = \partial_k\partial_l f_{\rm TAP}(m)$ is the 
Hessian
of the TAP free energy evaluated in a particular solution $m$. In what
follows the symbol ${\cal D}x$ refers to the integration over all the
site variables, ${\cal D} x = dx_1 \dots dx_N$.  Let us introduce the
following Grassmann representation for the determinant,
\beq
\det {\cal A}(m^a) =
	\int {\cal D} \bar\psi^a {\cal D} \psi^a 
	\exp \left[ - \bar\psi^a {\cal A}(m^a) \psi^a \right] ,   
	\label{fermi}
\eeq
and a bosonic representation for the delta functions, 
\beqa
\prod_{k=1}^N \delta \left( {\cal T}_k(m^a) \right) &=&  
	\int {\cal D} \lambda^a  \exp (- \lambda^a {\cal T}(m^a)) , \\
\delta \left( f_{\rm TAP}(m^a) - f \right) &=& 
	\int d \omega^a 
	\exp \left[ -\omega^a (f_{\rm TAP}(m^a) - f) \right] ,
	\label{del}
\eeqa
where the integrals over the variables $\lambda$ and $\omega$ are on
the imaginary axis. The sums over site indices are always understood
whenever site dependent quantities are mutually multiplied.  In the
rest of the calculation we shall disregard the modulus of the
determinant in (\ref{number}). This approximation is safe for zero
magnetic field, as long as we are counting stationary points with a
given {\it fixed} free energy density \cite{noiselle}. 
Indeed, it can be shown that in this
way only the minima of $f_{\rm TAP}$ are actually taken into
consideration \cite{noiselle}. We assume this to hold also with $h \neq
0$. We then have,
\beq
\Sigma(f) =
	\lim_{n \rightarrow 0} \frac{1}{n \, N} \log 
	\int {\cal D}m \, {\cal D}\lambda \, {\cal D}\bar\psi
	\, {\cal D}\psi \, d\omega \, 
	\overline{ 
	\exp \left[- {\cal S}_J(m,\lambda,\bar\psi,\psi,\omega) \right] },
	\label{sigma}
\eeq
where the action ${\cal S}_J$ is given by
\beq
{\cal S}_J = \sum_{a=1}^n \left[ \lambda^a {\cal T}(m^a) + 
	\bar\psi^a {\cal A}(m^a) \psi^a + 
	N \omega^a(f_{\rm TAP}(m^a)-f) \right] .
	\label{action}
\eeq
The average over the disorder generates couplings between variables
with different replica indices. It is convenient to express these
terms by means of the following overlap matrices, which can be
introduced in the usual way \cite{crisa1,crisatap,noiselle},
\beq
N S_{ab} = m^a m^b , \; \; \; 
N L_{ab} = \lambda^a \lambda^b , \; \; \;
N R_{ab} = m^a \lambda^b , \; \; \;
N T_{ab} = - \bar\psi^a \psi^b .
	\label{matrices}
\eeq
In terms of the overlap matrices (\ref{matrices}) we have,
\beq
\Sigma(f) = \lim_{n\rightarrow 0} \frac{1}{n N} \log 
	\int {\cal D}S \,  {\cal D}L \,   {\cal D}R \, {\cal D}T \, d\omega	
	\, \exp \left[ - N {\cal S}(S,L,R,T,\omega) \right] ,
	\label{integral}
\eeq
where the effective action is given by
\beqa
- {\cal S} &=& \frac{p}{4} \sum_{ab} L_{ab} S_{ab}^{p-1} 
	+ \frac{p(p-1)}{4} \sum_{ab} 
	\left( R_{ab}^2 - T_{ab}^2 \right) S_{ab}^{p-2}
	\non \\ 
	&& 
	+ \frac{h^2}{2} \sum_{ab} L_{ab}
	+ \sum_a 2 \left[ g'(S_{aa}) + \frac{1}{2\beta(1-S_{aa})}\right] 
		\left( T_{aa} - R_{aa} \right)
	- \log \det T 
	\non \\
	&&
	+ \frac{1}{2} \log \det S
	+ \frac{1}{2} \log \det \left( R^T S^{-1} R - L \right)
	+ \frac{p}{2} \omega \sum_{ab} R_{ab} S_{ab}^{p-1}
	+ h^2 \omega \sum_{ab}R_{ab}
	\non \\
	&&
	+ n f \omega - \omega \sum_a 
	\left[ g(S_{aa})-\frac{1}{2\beta}\log (1-S_{aa}) \right]	
	+ \frac{1}{4} \omega^2 \sum_{ab} S_{ab}^p
	+ \frac{h^2}{2} \omega^2 \sum_{ab} S_{ab} .
	\label{azia} 
\eeqa
We have assumed $\omega_a = \omega$ since it depends only on one
replica index. 
As usual, we compute the integral (\ref{integral}) by means of a saddle
point approximation, so that,
\beq
\Sigma(f) = \lim_{n\to 0} \frac{1}{n} {\rm Ext}_{\phi} \, 
	\left[ - {\cal S}(f,\phi) \right] ,
	\label{sigmalei}
\eeq
where $\phi$ stands for all the variational parameters, 
$S, L, R, T$ and $\omega$.
  
In order to simplify the saddle point equations for
the overlap matrices we make use of the fact that action
(\ref{action}) is invariant under the following
Becchi-Rouet-Stora-Tyutin (BRST) transformation
\cite{brst,parisisourlas,zinnjustin},
\beq
m^a_i \rightarrow m^a_i + \epsilon \, \psi^a_i 
	, \;\;\; 
	\bar\psi^a_i \rightarrow \bar\psi^a_i - \epsilon \, \lambda^a_i 
	, \;\;\; 
	\lambda^a_i \rightarrow \lambda^a_i 
	- \omega^a \, \epsilon \, \psi^a_i 
	, \;\;\;
	\psi^a_i  \rightarrow \psi^a_i 
	, \;\;\;
	\omega^a \rightarrow \omega^a 
	\label{trasform}
\eeq
where $\epsilon$ is a constant Grassmann parameter. As a consequence,
for each operator $O$, which is a 
function of the variables $m^a$, $\lambda^a$,
$\bar\psi^a$, $\psi^a$ and $\omega^a$, we have that $\langle \delta O
\rangle=0$, where $\delta O$ is the variation of $O$ under
(\ref{trasform}) and the brackets indicate an average over the measure
defined by the action ${\cal S}_J$. If we consider the two cases $O=m^b
\bar\psi^a$ and $O = \lambda^b \bar\psi^a$, we immediately get the
equations $\langle \bar\psi^a \psi^b \rangle= - \langle m^b
\lambda^a \rangle$ and $\langle \omega \bar\psi^a \psi^b
\rangle = \langle \lambda^a \lambda^b \rangle$.  From
the definitions (\ref{matrices}) we then obtain the following relations
for the saddle point values of the overlap matrices,
\beq
R_{ab} = T_{ab} , \;\;\;
	L_{ab} = -\omega \, T_{ab} 
	\label{final}
\eeq
which simplify a lot the calculation.  At this point we have to choose
an ansatz for the overlap matrices in order to solve the saddle point
equations.  Let us consider first of all the matrix $S$, which has an
explicit physical meaning. From the definition (\ref{matrices}) we see
that the diagonal element $S_{aa}$ corresponds to the self-overlap of
an individual TAP solution. On the other hand the off-diagonal elements
$S_{ab}$ correspond to mutual overlaps between different TAP solutions.
Given this we  assume for the matrix $S$ the simplest form 
consistent with this interpretation, that is a symmetric matrix,
\beq
S_{ab} = (s_1 - s_0) \delta_{ab} + s_0 .
	\label{ansatzs}
\eeq 
In the saddle point, $s_1$ and $s_0$ represent respectively  the
self-overlap and the mutual overlap of the states with free energy
density $f$.  As discussed above, in absence of the magnetic field
$s_0$ vanishes, the matrix $S$ is diagonal and the quenched average
coincides with the annealed one. On the other hand in the presence of
the field we expect a value $s_0 \neq 0$. Consistently with the ansatz
used for $S$ we set,
\beq
T_{ab} = (t_1 - t_0) \delta_{ab} + t_0 .
	\label{ansatz}
\eeq
Using the BRST relations (\ref{final}) and the above form for the
matrices $S$ and $T$, we can reduce the saddle point equations to the
following five coupled equations,
\beqa
0 &=& -\omega \left( \frac{p}{2} s_1^{p-1} + h^2 \right)
	- \frac{p (p-1)}{2} t_1 s_1^{p-2} 
	- \frac{1}{z} + \frac{1}{z^2} (\omega s_0+t_0)
	+ 2 g'(s_1)+\frac{1}{\beta(1-s_1)}  ,
	\non \\
0 &=& \omega \left( \frac{p}{2} s_0^{p-1} + h^2 \right)
	+ \frac{p (p-1)}{2} t_0 s_0^{p-2}
	- \frac{1}{z^2}(\omega s_0+t_0) ,
	\non \\
0 &=& - \omega \left( \frac{p}{2}s_1^{p-1} + h^2 \right)
	+ \frac{1}{t_1 - t_0} - \frac{t_0}{(t_1 - t_0)^2}
	-\frac{1}{z} + \frac{1}{z^2} (\omega s_0 + t_0) ,
	\non \\
0 &=& \omega \left( \frac{p}{2}s_0^{p-1} + h^2 \right)
	+ \frac{t_0}{(t_1 - t_0)^2}
	- \frac{1}{z^2} (\omega s_0 + t_0) ,
	\non \\
0 &=& \frac{p}{4} \left( t_1 s_1^{p-1} - t_0 s_0^{p-1} \right)
	+ \omega \left( \frac{1}{2} (s_1^{p-1} - s_0^{p-1})
	+ h^2 (s_1 - s_0) \right) ,
	\non \\
	&&
	- \frac{s_1}{2 z} 
	+ \frac{\omega s_0 + t_0}{2 z^2} (s_1 - s_0)
	+ f - g(s_1) + \frac{1}{2 \beta} \log (1 - s_1) ,
	\label{equations}
\eeqa
where $z= [\omega (s_1 - s_0) + t_1 - t_0]$ and $g(s)$ is the Onsager
term of (\ref{ftap}). It is easy to verify that for $h=0$ these
equations give the standard result of \cite{crisatap}.

\section{Connection with the statics}

We are now in the condition to compute the complexity $\Sigma$ in every
point of the plane $(T,h)$ and in particular in the regime where the
statics of the model display a 1-RSB solution (low temperatures and fields)
\cite{crisa1}. As we will show below, our results clarify the relation
between the present entropic approach and the usual static one.

First we note that equations (\ref{equations}) give, after some
algebra, the relations,
\beq
t_0 = 0 \;\;\; , \;\;\; t_1 = \beta (1 - s_1).
	\label{sol}
\eeq
The physical interpretation of these equations is very simple. The
saddle point value of the matrix $T_{ab}$ is related to the expectation
value $\langle \bar\psi^a \psi^b \rangle$ (see Eq. (\ref{matrices})) and
from (\ref{fermi}) it is clear that this expectation value is nothing
else than the average of the matrix ${\cal A}^{-1}$, which is by
definition the inverse of the Hessian of the TAP free energy.
Therefore, the parameter $t_1$ can be interpreted as the inverse
curvature of the free energy in a typical state with free energy $f$. On
the other hand, since $s_1$ is the self-overlap of this state, $\beta
(1 - s_1)$ is the the magnetic susceptibility $\chi$. In this way
equation (\ref{sol}) gives the expected static relation between the
fluctuation $t_1$ and the dissipation $\chi$ of an equilibrium system.
In other words, the size $(1-s_1)$ of a state has two different
contributions, a thermodynamic one proportional to the temperature and
a geometric one proportional to the inverse curvature of the mean field
free energy.  Equation (\ref{sol}) has not been recognized before in
the context of this calculation \cite{crisatap,kpz}, simply
because for $h=0$ the homogeneity of the $p$-spin model permits to
compute everything at zero temperature where $s_1=1$, so that the above
relation remains hidden.

Due to equation (\ref{sol}) we are left with only three parameters,
$s_1,s_0$ and $\omega$. As we have seen, the physical interpretation of
$s_1$ and $s_0$ is straightforward, while this is still not the case
for $\omega$. In order to better understand what is the role of
$\omega$ we note that in the saddle point we have (see equations 
(\ref{sigmalei}) and (\ref{azia})),
\beq
\frac{d \Sigma(f)}{d f} = \omega(f) \, .
\eeq
If we call $f_0$ the ground state free energy, defined by
$\Sigma(f_0)=0$, the number of states having extensive free energy
$Nf=Nf_0+\delta f$, is
\beq
{\cal N}(\delta f)\,  \sim e^{\Sigma'(f_0)\,  \delta f} = 
	\, e^{\omega(f_0)\,  \delta f} , 
	\;\;\;   
	\delta f\sim O(1) .
	\label{omex}
\eeq
This equation shows that $\omega(f_0)$ plays the same role as the
static parameter  $x$, which represents the breaking point in the 1-RSB
scheme \cite{ultra,mpv2,crisatap}.  Given this, it is clear that, for each
value of $(T,h)$, the 1-RSB computation and the present one 
must map one into the other with the following rule:
\beq
f_0 = f_{\rm RSB}  \;\; \Rightarrow  \;\; 
	s_1(f_0) = q_1 , \;\;\;
	s_0(f_0) = q_0 , \;\;\; 
\omega(f_0) = \beta x ,
	\label{mazurka}
\eeq
where $f_{\rm RSB}$ is the static 1-RSB equilibrium free energy and $q_1$,
$q_0$ and $x$ are the 1-RSB parameters of the static computation
(respectively, self-overlap, overlap and breaking point)
\cite{crisa1}.  It is possible to verify analytically that the
relations (\ref{mazurka}) are fulfilled by our saddle point equations:
the complexity is zero at the static free energy $f_{\rm RSB}$ 
and here the equations
for the parameters $s_1$, $s_0$ and $\omega$ reduce exactly to the
static equations  for $q_1$, $q_0$ and $\beta x$.

In this way we have shown that, despite the symmetric form of the
overlap matrices used in our calculation (see Eqs. (\ref{ansatzs}) and
(\ref{ansatz})), the complexity describes the statics
 of the model with the
same degree of accuracy as the 1-RSB calculation. This point has
always been rather mysterious, since it was not evident why the two
approaches should give the same results. Now the explanation is clear. 
At the ground state free energy density $f_0$ the two calculations
simply write in different ways the same quantities. In the RSB approach
the self-overlap $q_1$ of the states is introduced, while for the
complexity it is natural to deal with the curvature of the states,
because of the presence of the Hessian. Yet, these two quantities are,
as we have seen, trivially related.  Moreover, the breaking point $x$,
which is typical of the RSB calculations, has its counterpart in the
parameter $\omega$, which has a very simple interpretation in the
context of the complexity, being just the derivative of $\Sigma$ 
with respect to the free energy. 
Finally, the overlap matrix in the two cases has a
slightly different physical meaning and this is why different ansatz
are taken for them: in the RSB approach this matrix refers to the
overlap between {\it configurations}, while in the present context it
refers to overlap between {\it states}. Due to this, the present
approach has {\it one less step} of replica symmetry breaking, compared
with the standard static one.  Clearly, the price we have to pay for
having a symmetric overlap matrix is the knowledge of the explicit form
of the TAP free energy.  Interestingly enough, the parameter
$x$ drops from the symmetric overlap matrix, to reappear in disguise as
$\omega$. We believe this scenario to be true in each model exactly
solved by a 1-RSB static computation.

We stress that what is stated above only holds for $f=f_0$, where the
complexity is  zero. On the other hand, increasing $f$ provides
information on the whole spectrum of the states.

Finally, we would like to recall a completely different method by which
the complexity can be computed \cite{monasson,franzparisi}. 
This method, which is perhaps less transparent than 
the one presented in this paper,
relates the complexity $\Sigma(f)$ to the static free energy of
$r$ real replicas of the system in the following way: $\Sigma(r) =
\beta r^2 \frac{\partial F_{r}}{\partial r}$ and $f(r) = \frac{\partial
\left[ r F_{r} \right]}{\partial r}$, where $r F_r$ is the free energy
density of the $r$ coupled real replicas in the limit where the
coupling goes to zero.  To obtain $\Sigma$ it is therefore necessary to
compute $F_r$.  For systems with 1-RSB one usually assumes a one step of
breaking also for the overlap matrix related to $F_r$, with a fixed breaking
parameter equal to $r$ and two variational parameters $q_1$ and $q_0$
\cite{monasson,potters}. In this case the expression of $F_r$ as a
function of $q_1$, $q_0$ and $r$ is analogous to the usual static one,
except for the fact that it has to be minimized with respect to $q_0$
and $q_1$, but not with respect to $r$.  
With these assumptions it is possible to obtain
an explicit expression of $\Sigma$ in terms of $f$, $q_1$, $q_0$ and
$r$, where $q_1$ and $q_0$ are given by their saddle point values,
while $r$ is fixed by the above equation for $f(r)$.  If we now compare
this expression for $\Sigma$ with the one coming from 
our calculation,
we find that they are perfectly consistent: the equations for $q_1$,
$q_0$ and $r$ are exactly the same as the ones for $s_1$, $s_0$ and
$\beta \, \omega$, and $\Sigma(r)$ coincides with (\ref{sigmalei}).

\section{Behavior of the Complexity in the Temperature and Field
Plane}
	\label{sec:plane}

From equations (\ref{azia}), (\ref{sigmalei}) and (\ref{sol}) we obtain 
the explicit expression of the complexity 
\beqa
\Sigma(f) &=& \frac{p}{4} \omega t_1 s_1^{p-1} + 
		\frac{h^2}{2} \omega t_1 - \frac{1}{2} \log t_1
	 + \frac{1}{2} \log z + \frac{1}{2} \frac{\omega s_0}{z}
	\non 	\\
 	&&
	+ \omega \left( f - g(s_1) + \frac{1}{2\beta}\log (1-s_1)\right)
	+ \frac{\omega^2}{4} \left( s_1^p - s_0^p \right) 
		+ \frac{h^2}{2} \omega^2 (s_1-s_0) ,
	\label{pirla}
\eeqa 
with $t_1=\beta(1-s_1)$ and $s_1, s_0$ and $\omega$ given by the 
saddle point equations (\ref{equations}).

Let us now analyze the behavior of the complexity in the $(T,h)$ plane.
For small enough values of temperature and field many pure states
are present, therefore we expect the complexity to be different from
zero in a finite range of free energy densities. Actually this is what
happens. In Fig.  1 we show $\Sigma(f)$ as a function of $f$ at fixed
small values of $T$ and $h$: $\Sigma$ is defined between a minimum free
energy density $f_0$ where it is zero, and a threshold free energy
density  $f_{\rm th}$ where it takes its maximum value.  Since for high
values of temperature and field only one (finite magnetization) 
paramagnetic state is present \cite{crisa1}, we expect that 
increasing $T$ and $h$ the number of
existing states progressively decreases until one single state is left.
Indeed, if we look at $\Sigma(f)$ at a fixed $h$, but at
different values of $T$, we find that the interval of free energies
$[f_0,f_{\rm th}]$ where $\Sigma$ is defined, becomes smaller and smaller as
the temperature is increased, and finally it shrinks to a single point, with
$\Sigma=0$, at a certain critical temperature $T_c(h)$. For $T>T_c(h)$,
$\Sigma(f)$ is defined only in one point, where $\Sigma=0$, and $f$
here coincides with the replica symmetric static free energy, that is, the
free energy of the single paramagnetic state. In Fig. 2 the line
$T_c(h)$ is shown in the $(T,h)$ plane \cite{david}. 
This line separates the
region of the plane where only one pure state exists ($T>T_c(h)$) from
the region where many pure states are present  ($T<T_c(h)$)
and therefore we identify it as the {\em geometrical} transition line. 
We note that this line is monotonically decreasing with the field,
so that there is no `re-entrance' \cite{hertz}.
We find that the geometrical transition line
coincides with the line where the RSB static
solution ceases to exist \cite{crisa1}. 
This indicates that, as in the case with
$h=0$ \cite{kpz}, the states with the lowest free energy density are
the last to  disappear.  We note that the critical line $T_c(h)$ does
not coincide in the whole plane with the transition line given by the
statics (dotted line in Fig.  2) \cite{crisa1}.  Indeed the two lines
coincide only for $h>h^\star$, where the static transition is
continuous ($q_1=q_0$ at the transition), while they are strictly
different for $h<h^\star$ (discontinuous transition) \cite{crisa1}.
This is consistent with the different physical scenarios corresponding
to the continuous and the discontinuous transitions. In the first case,
the transition corresponds to different states which merge into a
unique ergodic component, so above the transition line only one state is
present. In the second case, many pure states
with finite complexity exist both above and below the transition line,
and the transition corresponds to the point where the lowest states
($f=f_0=f_{\rm RSB}$) become the relevant ones from a thermodynamical
point of view \cite{kpz,crisatap}.

\section{Threshold energy and dynamical behavior}
	\label{sec:dynamics}

As mentioned in the previous section, at fixed temperature and field,
the complexity is defined within a certain range $[f_0,f_{\rm th}]$ of free
energy densities.  It is well known that at $h=0$  in the low
temperature region the states with $f=f_{\rm th}$ are the dynamically
relevant ones, that is, their energy density coincides with the  asymptotic
value of the dynamical energy \cite{ck1,kpz}. In this section we show
that this also holds at non zero values of the field.  To check this
point we will perform a dynamical analysis of the model with $h\neq
0$ in order to compare the dynamical asymptotic energy with the energy
of the threshold states of the complexity. The equivalence between the
threshold and dynamical energies can be confirmed numerically for any
temperature and field, but for the sake of simplicity we will only
quote the analytical results at zero temperature and small field. 
If we take the limit $T \rightarrow 0$ of Eqs. (\ref{equations}) and 
expand it for
small fields we get the following expression for the
energy density of the threshold states, 
\beq
E_{\rm th}=	- \sqrt{\frac{2 (p-1)}{p}} 
	\left( 1 + \frac{(p-1)}{p} h^2 \right)
	+ O(h^3) , \;\;\;\; T\to 0. 
\label{thre}
\eeq

Let us now compute the asymptotic dynamical energy \cite{ck1}
(for a more complete dynamical analysis in similar cases see 
\cite{ledu,hertz}).  The relaxational dynamics for the
system is given by the Langevin equations,
\beq
\dot{s}_i(t) = - \beta \frac{\delta H}{\delta s_i(t)}
	- y(t) s_i(t) + \nu_i(t) ,
\eeq
where $y(t)$ is a Lagrange multiplier which enforces the spherical
constraint, and $\nu_i(t)$ is a Gaussian noise with zero mean and
variance 2. The dynamics is completely determined by
the equations for the two-time correlation and response functions,
$C(t,t') = \frac{1}{N} \sum_i \overline{\langle s_i(t)
s_i(t') \rangle}$ and $G(t,t') = \frac{1}{N} \sum_i
\overline{\frac{\partial \langle s_i(t) \rangle}{\partial h_i(t')}
}$,  and for the magnetization $m(t) = \frac{1}{N} \sum_i \overline{\langle
s_i(t) \rangle}$. These equations are \cite{ck1,sompozip},
\beqa
\partial_t C(t,t') &=&
	- y(t) C(t,t') + \mu \int_0^{t'} dt'' C^{p-1}(t,t'') G(t',t'') 
	\non \\
	&&
	+ \mu (p-1) \int_0^t dt'' G(t,t'') C^{p-2}(t,t'') C(t'',t') 
	+ \beta \, h \, m(t') ,
	\label{eqm0}
	\\
\partial_t G(t,t') &=&
	- y(t) G(t,t') + \delta(t - t')
	+ \mu (p-1) \int_{t'}^t dt'' G(t,t'') C^{p-2}(t,t'') G(t'',t') ,
	\\
\dot{m}(t) &=&
	- y(t) m(t)
	+ \mu (p-1) \int_0^t dt'' G(t,t'') C^{p-2}(t,t'') m(t'') 
	+ \beta \, h ,
	\label{eqm}
\eeqa
where $\mu = p \, \beta^2/2$. 
Another equation for $y(t)$ is obtained self-consistently exploiting 
the spherical constraint,
\beq
y(t) = 1 - p \, \beta \, {\cal E}(t) - (p-1) \beta \, h \, m(t) , 
\eeq
where ${\cal E}(t) =N^{-1} \overline{\langle H(t) \rangle}$ is the
dynamical energy density. 
In the limit $t \rightarrow \infty$ we have the following expression 
for the asymptotic dynamical energy 
${\cal E}_{\infty} = \lim_{t \rightarrow \infty} {\cal E}(t)$, 
\beq
{\cal E}_{\infty} = \frac{1 - y_{\infty}}{p \beta} 
	- \frac{p-1}{p} \, h \, m_{\infty} ,
	\label{ed} 
\eeq
where $y_{\infty}$ and $m_{\infty}$ are the corresponding asymptotic 
values for the Lagrange multiplier and the magnetization.
In order to compute ${\cal E}_{\infty}$ we then have to find $y_{\infty}$ 
and $m_{\infty}$ by solving asymptotically the set of dynamical
equations (\ref{eqm0})-(\ref{eqm}).

In \cite{ck1} it has been shown that for large times there are
two regimes for the correlation and response functions. The first
regime corresponds to time separations $\tau =(t-t')$ such that $\tau /
t \rightarrow 0$, where the equilibrium fluctuation dissipation theorem
(FDT) holds, so that $G_{\rm FDT}(\tau) = - \partial_{\tau} C_{\rm
FDT}(\tau)$. The second regime, known as aging regime, corresponds to
$\tau/t \sim O(1)$. Here time translation invariance is violated 
and FDT cannot be applied. The response function is
related to the correlation function by $G_{\rm ag}(t,t') = x_d
\, \partial_{t'} C_{\rm ag}(t,t')$, where $x_d$ parameterizes the
violation of FDT. The asymptotic values for the correlation functions
in both regimes are,
\beq
\lim_{\tau \rightarrow \infty} C_{\rm FDT}(\tau) = 
	\lim_{t'/t \rightarrow 1} C_{\rm ag}(t, t') = 
	q , \;\;\;
\lim_{t'/t \rightarrow 0} C_{\rm ag}(t, t') = q_0 .
\eeq
In the limit $\tau \rightarrow \infty$, the equation for $C_{\rm
FDT}(\tau)$ yields a relation between $y_{\infty}$ and $q$ which reads,
\beq
y_{\infty} = (1 - q)^{-1} + \mu (1 - q^{p-1}) \ .
\eeq
If we take the limit $t'/t \rightarrow 1$ in the equation for $G_{\rm
ag}(t, t')$ we obtain the equation for $q$,
\beq
\mu q^{p-2} (1 - q)^2 - (p - 1)^{-1} = 0 .
	\label{eqq}
\eeq
It is now clear that neither $q$ nor $y_{\infty}$ depend on the field. 
Therefore, all the dependence of the energy on the field is given by the
second term of Eq. (\ref{ed}), while the first term corresponds to the 
asymptotic energy in zero field \cite{ck1}.

To obtain the asymptotic magnetization $m_{\infty}$ we need to solve
the coupled equations for $m_{\infty}$, $q_0$ and $x_d$ coming from Eq.
(\ref{eqm}), and from the limits $t'/t \rightarrow 0$ and $t'/t
\rightarrow 1$ of the equation for $C_{\rm ag}(t, t')$,
\beqa
0 &=& - y_{\infty} m_{\infty} 
	+ \mu \left[ 1 - q^{p-1} (1 - x) \right] m_{\infty} 
	+ \beta \, h ,	
	\\
0 &=& - y_{\infty} q_0 
	+ \mu (1 - q) q_0^{p-1} + \mu q_0 (1 - q^{p-1})
	+ \mu x_d (q^{p-1} - q_0^{p-1}) + \beta \, h \, m_{\infty} ,	
	\\
0 &=& - y_{\infty} + 1 + \mu (x - 1) q^p + \mu (1 - x q_0^p) 
	+ \beta \, h \, m_{\infty} .
\eeqa
The analytic solution for small $h$ has a simple form,
\beqa
m_{\infty} &=& (p-1) (1-q) \, \beta \, h + O(h^2) , \\
q_0 &=& O(h^2) , \\
x_d &=& \frac{(p-2)(1-q)}{q} + O(h^2) .
\eeqa
From here we obtain the dynamical energy,
\beq
{\cal E}_{\infty} = - \frac{\beta}{2} \left[ 1
	- q^p \left( 1 - \frac{(p-2)(1-q)}{q} \right) \right] 	
	- \frac{(p-1)^2 (1-q)}{p} h^2 + O(h^3) . 
	\label{suka}
\eeq
In the limit $T \rightarrow 0$ it can be seen from equations (\ref{eqq}), 
(\ref{suka}) and (\ref{thre}) that ${\cal E}_{\infty}=E_{\rm th}$.
As already said, it is possible to check numerically that these two
energies are the same in the whole plane $(T,h)$.

\section{Conclusions}
	\label{sec:conclusions}

In this paper we have computed the complexity $\Sigma$ of the states in
a very general 1-RSB mean-field model for disordered systems, namely,
the $p$-spin spherical model in presence of an external magnetic field.

We find a region of the $(T,h)$ plane where $\Sigma$ is a  monotonic
increasing function of the free energy density $f$, defined in an
interval $[f_0,f_{\rm th}]$. 
The lower band edge
$f_0$, defined by $\Sigma(f_0)=0$, is always equal to the free energy
density of the static 1-RSB solution, that is, $f_0=f_{\rm RSB}$.  On the
other hand, the threshold value $f_{\rm th}$ gives the free energy density
of the states asymptotically reached by the non-equilibrium dynamical
evolution of the system. 

Our most important result is the one to
one mapping we are able to find between the equations and the
parameters of the complexity at $f=f_0$ and the ones of the 1-RSB static
approach.  This shows that the two calculations are essentially the
same, even if the ansatz taken for the overlap matrices seem so
different in the two cases. 

We would like to add some further comments
on what is the main interest of the present work.
There are different methods by which the complexity of a model can be
computed \cite{braymoore,franzparisi,monasson}. The one used in this
paper is the most intuitive and transparent, since it simply counts 
the number of minima of the mean-field free energy function.
However, from a technical point of view, this method is far from
being straightforward and up to now a clear quenched calculation of the 
complexity was lacking. Many efforts have been done in this direction
for the Sherrington-Kirkpatrick model \cite{sk}, 
which has a full RSB static solution 
\cite{innocent1,innocent2,brayfull,potters}.
However, in that case neither the physical meaning of the adopted ansatz
nor the eventual consistency of the lower band edge with the static
free energy density were completely satisfying. 
In the present paper, in the context of a simpler 1-RSB model, we find in
a clear way how the computation has to be performed and what is the meaning
of the ansatz. Therefore, we now believe to have this method
under control.

It would be interesting to use this same method for a deeper
analysis of the free energy landscape. Indeed, it is generally believed
that in the dynamics of glassy systems a crucial role is
played not only by the minima of the Hamiltonian, but also by 
unstable stationary points \cite{laloux,noiselle}. 
In this context, an entropic computation of stationary points of 
any nature is of primary interest.

\acknowledgements

We thank Andrea Crisanti, Leticia Cugliandolo, John Hertz, and especially 
Jorge Kurchan and Giorgio Parisi for helpful discussions.  We also
thank David Sherrington for useful suggestions and for carefully
reading the manuscript. The work of A.C.  and I.G was supported by
EPSRC Grant GR/K97783. The work of J.P.G. was supported by EC Grant
ARG/B7-3011/94/27.

\pagebreak

\begin{figure}
\begin{center}
\leavevmode
\epsfxsize=6in
\epsffile{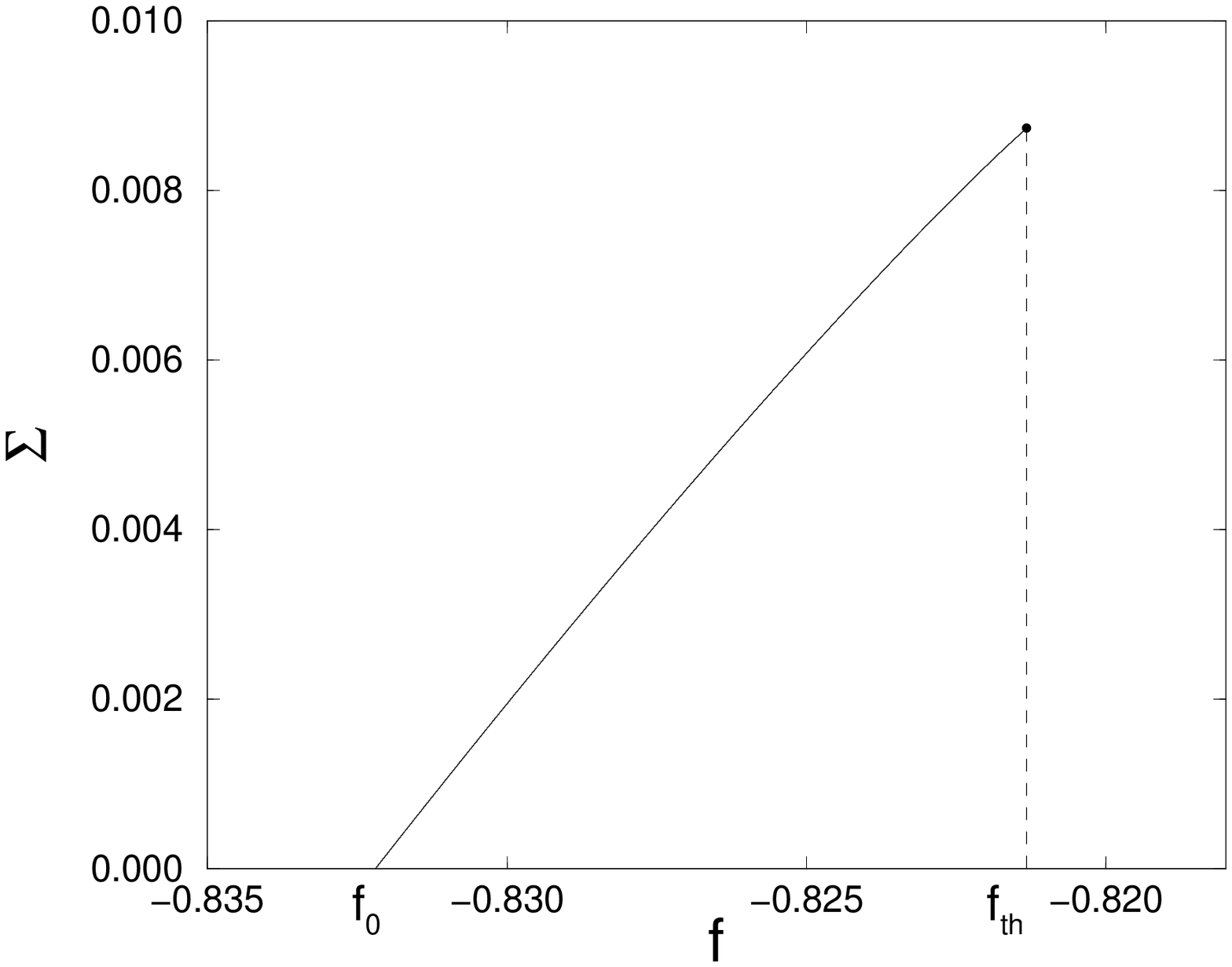}
\caption{The complexity $\Sigma(f)$ as a function of the free energy
density $f$ for $T = 0.2$, $h = 0.2$ and $p = 3$. The minimum value
$\Sigma = 0$ occurs at $f_0 = f_{\rm RSB} = -0.8322$. The maximum value
is at $f_{\rm th} = -0.8213$, which is the free energy density of the
states reached by the dynamics.}
\label{fig1}
\end{center}
\end{figure}

\begin{figure}
\begin{center}
\leavevmode
\epsfxsize=6in
\epsffile{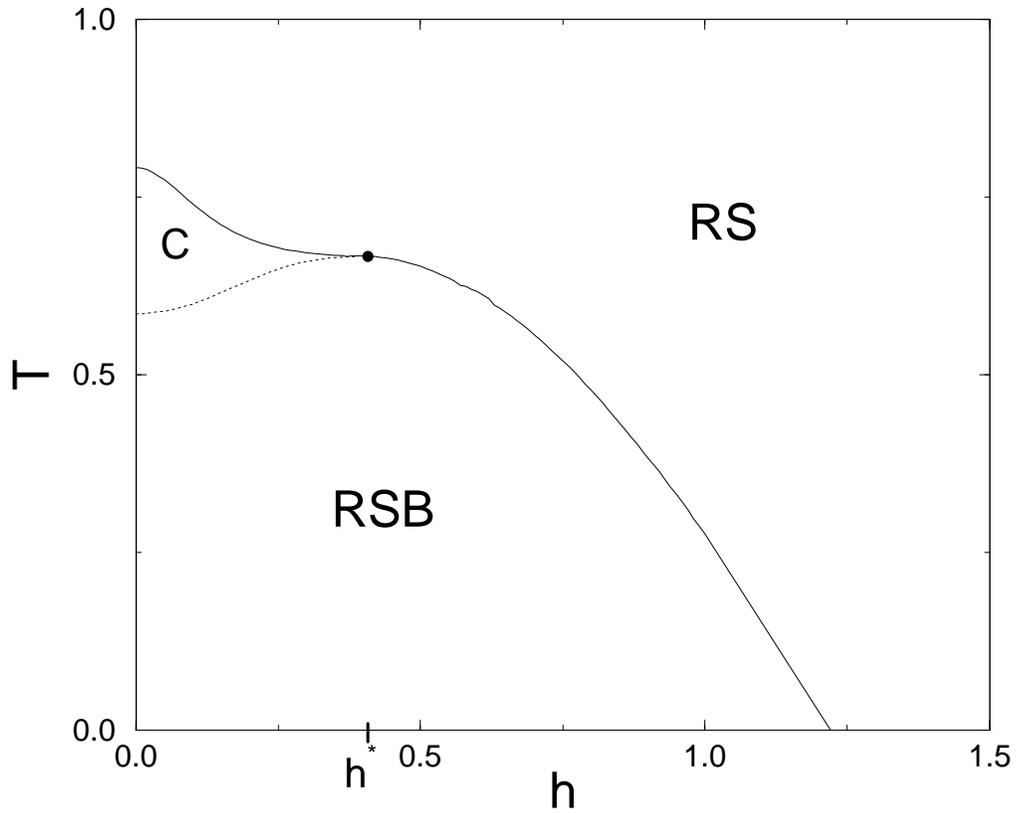}
\caption{The full line corresponds to the critical temperature line
$T_c(h)$ where $\Sigma$ reduces to a point and the RSB static solution
disappears, for $p = 3$. The dotted line corresponds to the static
transition line. The two lines are coincident for fields $h > h^\star =
0.408$. The static transition is continous for $h > h^\star$, and
dicontinous for $h < h^\star$. The region C is the region of coexistence of
RS and RSB solutions.}
\label{fig2}
\end{center}
\end{figure}

\end{document}